\documentclass[letterpaper]{article}
\usepackage{iccc}
\usepackage{verbatim}
\usepackage{multirow}
\usepackage{graphicx, array, blindtext}
\usepackage{capt-of}
\usepackage{tabu}
\usepackage{times}
\usepackage{dirtytalk}
\usepackage{helvet}
\usepackage[T1]{fontenc}
\usepackage{courier}
\usepackage{xurl}
\usepackage{natbib}
\title{Exploring Real-Time Music-to-Image Systems for Creative Inspiration in Music Creation}
 \author{Meng Yang, Maria Teresa Llano, Jon McCormack \\
 Sensilab\\
 Faculty of IT, Monash University\\
 \{Meng.Yang, Teresa.Llano, Jon.McCormack\}@monash.edu}

\begin{document} 
\maketitle
\begin{abstract}

This paper presents a study on the use of a real-time music-to-image system as a mechanism to support and inspire musicians during their creative process. The system takes MIDI messages from a keyboard as input which are then interpreted and analysed using state-of-the-art generative AI models. Based on the perceived emotion and music structure, the system’s interpretation is converted into visual imagery that is presented in real-time to musicians. We conducted a user study in which musicians improvised and composed using the system. Our findings show that most musicians found the generated images were a novel mechanism when playing, evidencing the potential of music-to-image systems to inspire and enhance their creative process.

\end{abstract}

\section{Introduction}

Artificial Intelligence (AI) has achieved remarkable results in music generation and analysis \citep{deepBach,MusicTransformer,jukebox,DeepComposer}, as well as in its applications to co-creation with human musicians \citep{Robotic,TempoImprovision,co-creation_generative}. The conventional paradigm of human-AI music co-creation has predominantly revolved around AI-generated musical responses, which effectively eases the process of music creation and lowers the barriers to creative engagement. However, these advanced models and the refined AI-generated musical outputs present some challenges, with users finding themselves grappling with uncontrollable and unpredictable musical outputs that may not always aligned with the user's musical goals and intentions \citep{co-creation_generative}. 
Deviating from this well-trodden path, our work focuses on creating systems for music inspiration; thus, rather than generating uncontrollable musical results, as conventional methods have done, we employ visual stimuli associated with the perceived emotions of music as a source of inspiration. By leaving the primary work of music creation to the users, our system inspires them in a non-musical and subjective manner, fostering a cross-modal creative experience.

In this work, we delve into the innovative application of state-of-the-art generative AI models, aimed at providing inspiration and enhancing creativity through real-time responses to human musicians. In particular, we explore the use of Large Language Models (LLMs) for music interpretation. Although these models are mainly designed for natural language, their applications extend beyond texts and reach the field of music \citep{musilingo,GPT-2_evaluation}. Music is essentially a series of notes and rhythms, which can be seen as another form of language for models to learn, understand and generate. The powerful pattern recognition capability and context-level understanding of large language models may suggest new perspectives for the application of AI in music.

Our system consists of three components: A MIDI keyboard, a GPT-based music analyser for music analysis and prompt generation, and an image generator based on Stable Diffusion. When the user plays the keyboard, the music played is fed into the system as real-time MIDI information and is converted into a text-based music format for GPT-4 to analyse, which then generates a description of the visual imagery evoked by music based on the perceived emotion and musical structural features. This descriptive output serves as the prompt to generate the corresponding visual image, which is displayed to the user, forming a real-time music-to-image system. 

To explore the utility of the system in inspiring creativity in music creation, we conducted a user study with five participants. They engaged with the system under two music creation scenarios: improvisation and composition. To guide the study design and analysis, we proposed three research questions: 
 \begin{itemize}
  \item \textbf{RQ1}: How do images generated based on musical input affect the music-creation process of musicians?
  \item \textbf{RQ2}: What is the impact on musicians' creative processes when employing a divergent versus a convergent approach for the generation of images?
  \item \textbf{RQ3}: What factors can contribute to enhancing the experience of a music-to-image system?
\end{itemize}

The results from the study highlight the system's notable utility in improvisation, demonstrating its promise as a valuable inspirational tool for creation.

\section{Background}
\subsection{Visual Stimulation in Music Creation}
Music transcends mere auditory sensations, often evoking vivid visual imagery in listeners. For those with synesthetic abilities, for example, music manifests specific colours, while others might associate it with abstract forms \citep{Music-shape}. Some people, deeply moved by music, find themselves immersed in a visual narrative triggered by the music \citep{narrative}. This interplay between sound and sight significantly influences a musician's creative process. Composers often use mental visualisations of scenes or visual imagery involving elements of the physical world and `translate' these visual cues into sonic expressions. A study has demonstrated the positive impact that visual mental imagery can have on musical composition creativity through an experiment \citep{image_creativity}. Due to the rich perceptual experience of images, this cross-modal approach provides a broad exploration of musical concepts and therefore could be a valuable source of inspiration within the creative process. 

Previous studies have explored generating visual imagery from musical inputs. For example, one system uses a genetic algorithm to create abstract images inspired by music, where melodies are linked to the imagery's foreground, while harmonies are correlated with the background \citep{generic}. This approach is innovative, but may not serve as the most effective source of inspiration. Research shows that abstract images make up only a small portion of the music-relevant visual imagery reported by participants \citep{narrative}. The majority of reported visual imagery is tied to more realistic aspects, including narratives, environments, and characters, suggesting that a diverse spectrum of visual imagery could elicit a deeper emotional engagement. 

Our research does not impose any limitations on image styles, and images are generated based on perceived emotions from music, aiming to enhance the creator's internal feelings and create an immersive atmosphere.

\subsection{Generative Pre-trained Transformer}
Generative Pre-trained Transformers, commonly known as GPT, are a type of transformer-architecture Deep Learning model that brings breakthroughs in generative AI. Beyond their capability to generate high-quality human-like output, GPT's application in the realm of music has garnered increasing attention among researchers. One well-known example is MuseNet \citep{MuseNet}, developed by Open AI, which fine-tuned GPT-2 to generate musical compositions across ten different instruments and a variety of styles. The systematic evaluation of GPT-2's musical outputs, as discussed in \cite{co-creation_generative}, underscores the model's potential in this domain. Generating text-based representations of music (e.g., ABC notation) has also been attempted, where researchers re-trained  GPT-2 with a large dataset in ABC notation to generate folk music \citep{interactive_gpt2}. 

However, for GPT-3 and subsequent versions, research has been more limited due to their closed-source nature and restricted access. A test was conducted to evaluate the ability of GPT-3 to assist with songwriting and music production, particularly on lyrics creation and song recommendation tasks based on the time signature, instruments, sound physics and musical production, suggesting that GPT-3 is capable of enriching the creative process \citep{GPT_assistant_new}. Similarly, another research assessed GPT-3's capability in musical tasks such as identifying key signatures in given musical sequences and rationalizing musical decisions \citep{Toward_GPT3}. Despite these studies revealing GPT-3's current limitations in fully comprehending and processing complex musical tasks, they offer insights into the model's prospective applications in music.

These works focus on the objective understanding and analysis of music, explored via musical features, structures, and theoretical underpinnings.  The subjective dimensions of music, such as emotion, cultural/social context, association and cognitive interpretation, have seldom been addressed in the applications of Large Language Models within music. This study aims to delve into GPT's capacity to subjectively interpret and analyze music, exploring how it can link text-based musical elements with music's subjective attributes and descriptions.

\begin{table*}[!ht]

\centering
\small
\setlength{\tabcolsep}{1.2pt}
\renewcommand{\arraystretch}{1.7}

\begin{tabular}{{p{0.42\columnwidth}|p{1.58\columnwidth}}}
\multicolumn{1}{c|}{Features}                            & \multicolumn{1}{c}{Instrument}                            \\ \hline\hline
Instrument                          & The piece specifies two voices (V:1 and V:2) with the same program number (Program 1 0), indicating they are both intended to be played on the piano. The second voice is labeled as "bass,"  suggesting it plays a lower, supportive role in the harmonic texture.                                                     \\
Time Signature                      & The M:4/4 marking indicates a time signature of 4/4.                                                                                                                                                                                                                                                                    \\
Note Length                         & The L:1/8 setting specifies that the default note length is an eighth note (quaver).                                                                                                                                                                                                                                    \\
Tempo                               & The Q:1/4=96 setting indicates a tempo of 96 quarter notes (crotchets) per minute.                                                                                                                                                                                                                                      \\
Key Signature                       & The K:Em setting indicates the piece is in E minor.                                                                                                                                                                                                                                                                     \\ \hline
\multirow{2}{*}{Melody and Harmony} & The melody line in V:1 features a repetitive motif with variations in pitch, primarily using longer note values (mostly dotted half notes followed by eighth notes), which indicates a melodic theme that evolves gradually.                                                                                            \\ \cline{2-2} & The bass line in V:2 provides the harmonic foundation with a series of chordal patterns that repeat and evolve, anchoring the piece in its E minor tonality. The use of brackets indicates chords played together, with the progression moving through various harmonic intervals and contributing to the overall mood. \\ \hline
Dynamics and Articulation           & The use of "\textgreater{}" (accent) in V:1 suggests moments of emphasis or articulative interest, particularly in shaping the melody line.                                                                                                                                                                             \\
Repetition and Variation            & Both voices exhibit repetition and variation—a common compositional technique  to create cohesion and contrast. Voice 1 repeats a rhythmic pattern with slight melodic variations, while Voice 2 repeats a harmonic pattern with changes that reflect the melodic alterations in Voice 1.                              
\end{tabular}

\caption{The result of GPT-4 extracting musical features from input music in ABC Notation using the prompt "What musical features can you extract from the following musical piece written in ABC Notation? \{\textit{Music in ABC Notation}\}"}\label{tab1}
\end{table*}

\begin{table*}[!ht]
\centering
\small
\setlength{\tabcolsep}{4.3pt}
\renewcommand{\arraystretch}{1.7}
\begin{tabular}{{p{0.4\columnwidth}|p{1.6\columnwidth}}}
\multicolumn{1}{c|}{Features}                & \multicolumn{1}{c}{Inference}                                                                                                                                                                                                                                     \\ \hline\hline
Key Signature  & The choice of E minor sets a somber or melancholic tone for the piece. Minor keys are generally associated with sadness, introspection, or a contemplative mood.                                                                                                  \\
Tempo      & The moderate tempo of 96 beats per minute supports a reflective and somewhat introspective atmosphere. It’s just right for a contemplative or pensive mood.                                                                                                       \\
Melodic Repetition      & The melody in the V:1 line features repeated phrases with slight variations. This repetition can create a sense of longing or unresolved emotion, contributing to a reflective or melancholic atmosphere.                                                         \\
Harmonic Support        & The V:2 line provides a harmonic foundation that complements the melody. The chords and bass notes in the harmony reinforce the key’s emotional character, adding depth and richness to the piece’s overall mood.                                                 \\
Dynamic Contrast        & The potential for dynamic contrast between the melody and harmony lines, and within the melodic line itself, further emphasize the piece’s emotional depth. The interplay between louder and softer passages can enhance feelings of introspection or melancholy.
\end{tabular}
\caption{The result of GPT-4 inferring emotion from musical features of input music in ABC Notation using the prompt "What emotion can you infer from this musical piece written in ABC Notation? \{\textit{Music in ABC Notation}\}"}\label{tab2}
\end{table*}

\subsection{Human-AI Collaboration in Music}
The intersection of AI and human creativity, particularly in the realm of music, has garnered increasing interest with the advent of AI technology. The development of models for music generation, such as Google's Magenta \citep{Magenta}, OpenAI's MuseNet and Jukebox \citep{MuseNet,jukebox}, have significantly enhanced the calibre of AI-generated music, marking notable advancements in the field. However, the inherent unpredictability and lack of controllability in these models make them less ideal for musicians seeking a collaborative partnership with AI.  A study by \cite{control} indicates that artists expect more initiative and control when working with AI systems.

Therefore, research on  AI as an assistive creative tool for musicians during the music creation process becomes crucial. Tasks such as generating suitable chord progressions from provided melodies \citep{chord_generator1,chord_generator2}, completing melodies from specified notes \citep{jukebox,melody_generation}, generating variation of melodies \citep{variation}, and automatic accompaniment \citep{accomp1,accomp2}  are vital areas of focus. To deepen musicians' engagement in the co-creative process, innovative interaction modalities have been developed. For instance, in \cite{Gesture} a system was developed to enable musicians to steer the co-creation process through speech and gesture, facilitating expressive performances. In another study, communication was conducted through a designed conversational interface, where a Chatbot was created to understand the musicians' compositional intentions, which was then used to guide music composition using multiple algorithms through cross-modal interaction \citep{interface}. These methods allow musicians to communicate with the system through extra-musical interactions, and enable the system to adjust its output in response to user feedback, thereby enhancing more dynamic and collaborative creation processes.

Despite these innovations, the majority of studies have focused on employing AI-generated musical output as an assistant tool for music creation, while the influence of AI provided extra-musical inspiration on the creative process has remained largely unexplored.

\section{Musical Ability of GPT}
Despite being initially designed for natural language processing, the robust pattern recognition and generative capabilities of GPT models exhibit potential for application in music. In this work, we use \texttt{gpt-4-0125-preview}, the most recent iteration of the GPT-4 Turbo preview model, to explore its ability to understand text-based music. Our work is conducted with musical inputs presented as ABC Notation, a concise, computer-readable format that translates music into plain text. Its simplicity and efficiency in representing musical elements make it the ideal choice for inputting music to the GPT model.

\subsection{Extracting Musical Features from the ABC Notation}
To test the ability of the GPT-4 model to read text-based music, we used ABC Notation as input and asked what musical features it could extract from a given ABC Notation piece. The example we showed comes from the first 12 bars of Chopin's \textit{Prelude in E Minor (Op. 28, No. 4)}. Explicit information such as the name of the input piece was removed from the header of ABC Notation to ensure that the features extracted by GPT-4 come only from the text-based musical input.

\begin{table*}[!ht]
\centering
\small
\setlength{\tabcolsep}{25pt}
\renewcommand{\arraystretch}{1.5}
\begin{tabular}{lll}
\multicolumn{1}{c}{Musical piece}                 & \multicolumn{1}{c}{Genre} & \multicolumn{1}{c}{Emotion Inferred by GPT-4} \\ \hline \hline
Rondo Alla Turca (Turkish March) - Mozart         & Classical                 & Energetic, Lively, Complex                    \\
Sonate No. 14, Moonlight 3rd Movement - Beethoven & Classical                 & Vibrant, Intense, Energetic                   \\
Clair de Lune - Debussy                           & Classical                 & Melancholic, Reflective, Serene               \\
Nocturne Op. 48 No.1 - Chopin                     & Classical                 & Reflective, Tranquil, Introspective           \\
The Piano Medley - Queen                          & Pop                       & Dynamic, Bright, Whimsical                    \\
Quantifiable connection - Interstellar            & Pop                       & Urgent, Spirited, Exciting                    \\
Feel My Pain - Jurrivh                            & Pop                       & Meditative, Subdued, Somber                   \\
La Storia - Alexander Descartes                   & Pop                       & Tranquil, Harmonious, Soothing                \\
Don’t Jazz Me Rag - James Scott                   & Jazz                      & Joyful, Lively, Upbeat                        \\
Suite Dansante en Jazz - Schulhoff                & Jazz                      & Serene, Elegant, Graceful                     \\
London Blues - Jelly Roll Morton                  & Jazz                      & Expressive, Uplifting, Bright                
\end{tabular}
\caption{The result of GPT-4 inferring emotion of different musical pieces using the prompt "What emotion can you infer from this musical piece written in ABC Notation? Answer this question using only three words. \{\textit{Music in ABC Notation}\}"}\label{tab3}
\end{table*}
Table \ref{tab1} shows the related musical features extracted from the input ABC Notation. The results show that GPT-4 is able to obtain correct and reliable features from text-based music. While many of these features come from the header information in ABC Notation directly, some are inferred from musical patterns and structure.

\begin{figure}
\centering
\includegraphics[width=0.4\textwidth]{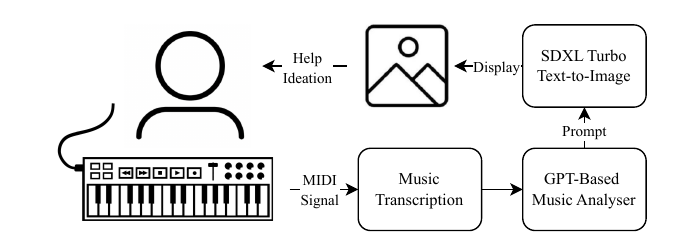}
\caption{The architecture and workflow of the system.}
\vspace{-0.3cm}\label{fig1}
\end{figure}

\subsection{Text-based Music Emotion Recognition through GPT-4}
Further experiments were conducted aiming to test whether GPT-4 can infer the emotion, a non-intuitively subjective feature, from text-based music representations. Despite the subjective nature of emotion perception in music, as stated by \cite{Gabrielsson}, two different emotional states should be distinguished. One is the perceived emotion, which is the listener's perception of the emotions conveyed in the music, a perceptual-cognitive process that arises from the decoding of the music, and another is the induced emotion, the emotional response evoked by music, which depends on personal experiences, current mood, etc. For the purpose of testing GPT-4, the focus is on perceived emotion, which is akin to existing Music Emotion Recognition tasks. The results in Table \ref{tab2} demonstrate that GPT-4 appears to be capable of inferring perceived emotions from intuitive musical features like key, time signature, tempo, rhythm, and melodic patterns.

For the needs of this project, we tested musical pieces from different genres that are suitable for playing solo on the keyboard, like Classical, Pop and Jazz, to test GPT-4's ability of emotion recognition in input musical pieces. For this iteration, the inference process in the result was required to be omitted and the output would be only three words to describe the emotion inferred from music.
Results in Table \ref{tab3} indicate that GPT-4 has a certain ability to recognize more diverse and specific emotion in these classical music examples. However, for Jazz examples, the results are more generic.

Overall, GPT-4 has a basic understanding of text-based music format in ABC Notation and is able to extract intuitive musical features and infer some deeper information based on these features. Although results are overall general and not completely reliable, for the purposes of our system, GPT-4 provides a good source to perform automatic interpretations of musical pieces written in ABC notation. 

\section{Real-time Music-to-Image Co-creation System}
Based on GPT-4's basic understanding of ABC Notation, we developed a real-time Music-to-Image co-creation system, where GPT-4's interpretation served as the input prompt to the text-to-image system, in this case, Stable Diffusion.

\begin{table*}[!ht]
\centering
\resizebox{1\linewidth}{!}{
\begin{tabular}{lll}
Music &
   \multicolumn{1}{c}{Prompt Generated By GPT-4} &
  \multicolumn{1}{c}{Visual Imagery} \\ \hline\hline
Clair de Lune - Debussy &
  \begin{tabular}[c]{p{1\columnwidth}}This music paints a picture of a serene, moonlit landscape, with gentle waves lapping at the shore of a tranquil lake, surrounded by silhouettes of towering mountains under a starry sky, evoking a sense of peace and quiet introspection.\end{tabular} &
\begin{minipage}[b]{0.35\columnwidth}
		\centering
		\raisebox{-.5\height}{\includegraphics[width=\linewidth]{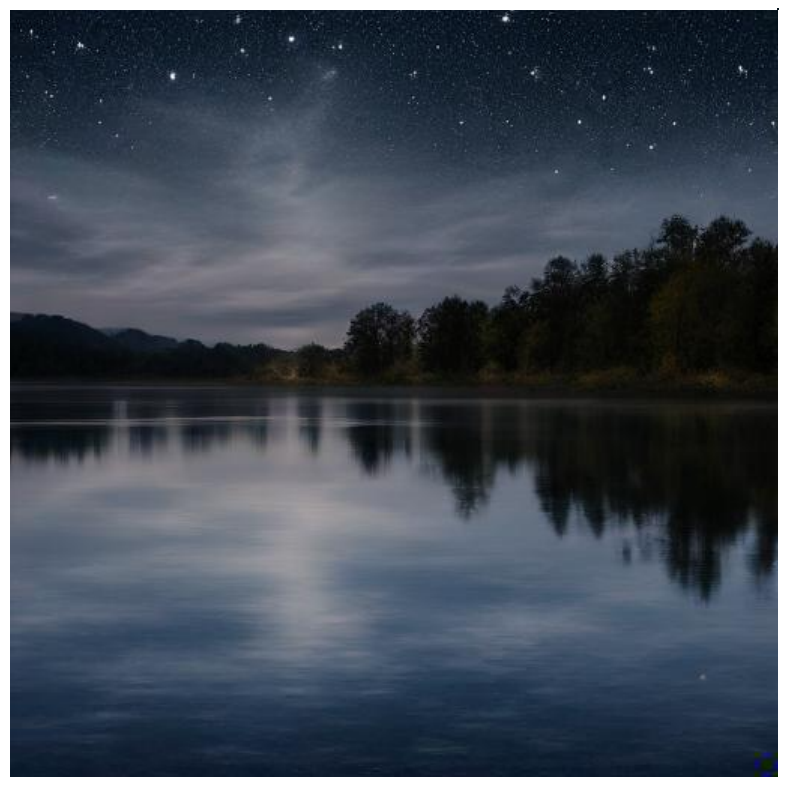}}
	\end{minipage}
  
   \\
\begin{tabular}[c]{@{}l@{}} Sonate No. 14, Moonlight \\ 3rd Movement -Beethoven\end{tabular}&
  \begin{tabular}[c]{p{1\columnwidth}}The music evokes a sense of dramatic urgency, likely belonging to the classical genre. Visualize a stormy sea at night, waves crashing against a lone, illuminated lighthouse on a cliff.\end{tabular} &
  \begin{minipage}[b]{0.35\columnwidth}
		\centering
		\raisebox{-.5\height}{\includegraphics[width=\linewidth]{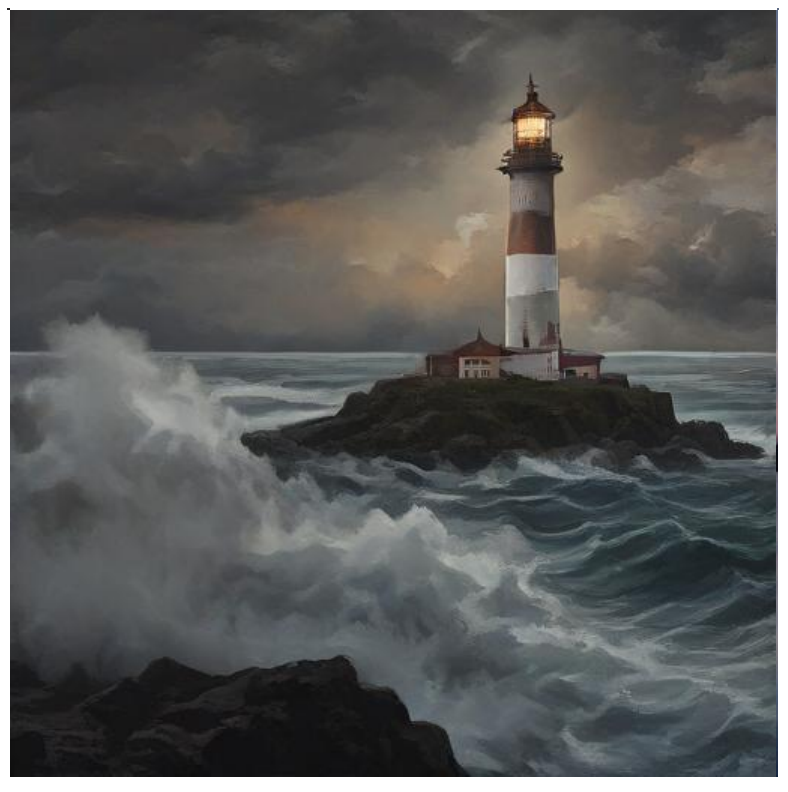}}
	\end{minipage}
   \\ 
Nocturne Op. 48 No.1 - Chopin &
  \begin{tabular}[c]{p{1\columnwidth}}A dimly lit, cozy room with rain tapping gently against the window, shows a sense of melancholic nostalgia.\end{tabular} &
    \begin{minipage}[b]{0.35\columnwidth}
		\centering
		\raisebox{-.5\height}{\includegraphics[width=\linewidth]{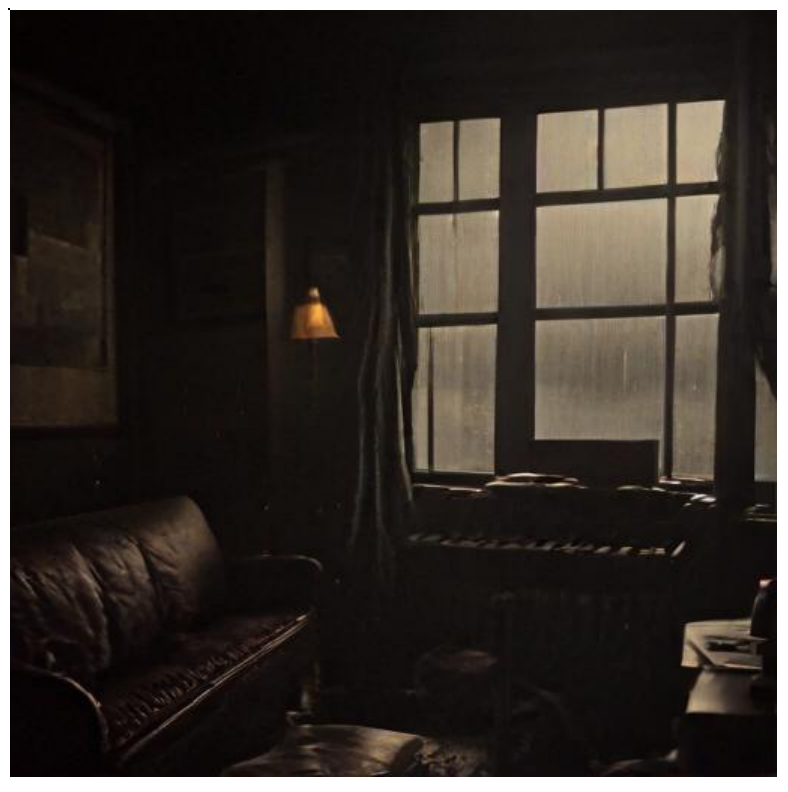}}
	\end{minipage}
   \\
Don’t Jazz Me Rag - James Scott &
  \begin{tabular}[c]{p{1\columnwidth}}A lively jazz club from the 1920s, with dancers in flapper dresses, a bustling crowd, and musicians lost in their energetic performance  under dim, flickering lights.\end{tabular} & 
      \begin{minipage}[b]{0.35\columnwidth}
		\centering
		\raisebox{-.5\height}{\includegraphics[width=\linewidth]{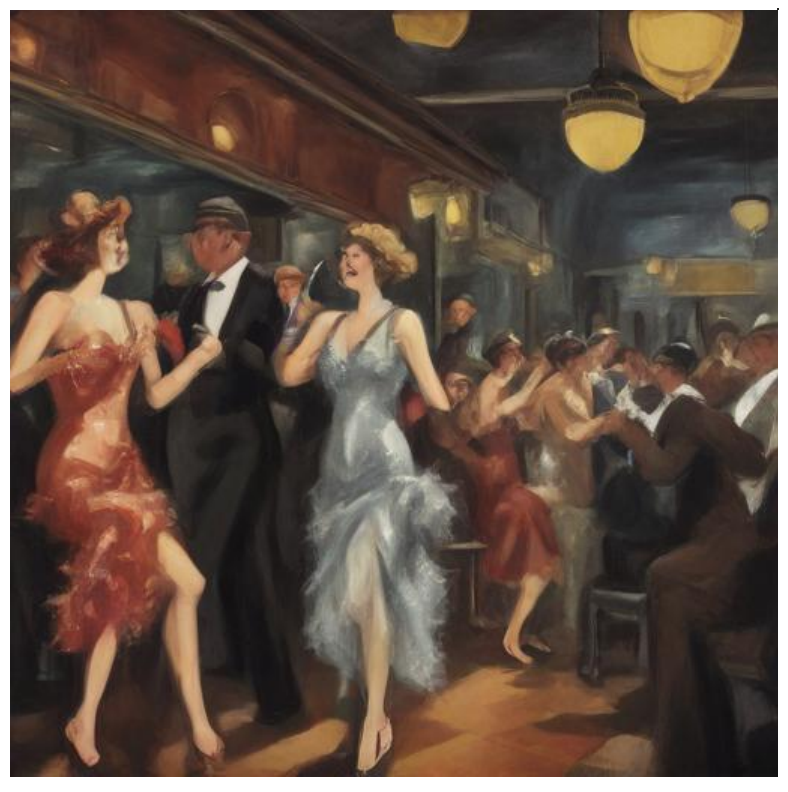}}
	\end{minipage}

   \\
The Piano Medley - Queen &
  \begin{tabular}[c]{p{1\columnwidth}}A moonlit dance in an ancient, enchanted forest, with leaves swirling in a gentle breeze under twinkling stars.\end{tabular} &
  \begin{minipage}[b]{0.35\columnwidth}
		\centering
		\raisebox{-.5\height}{\includegraphics[width=\linewidth]{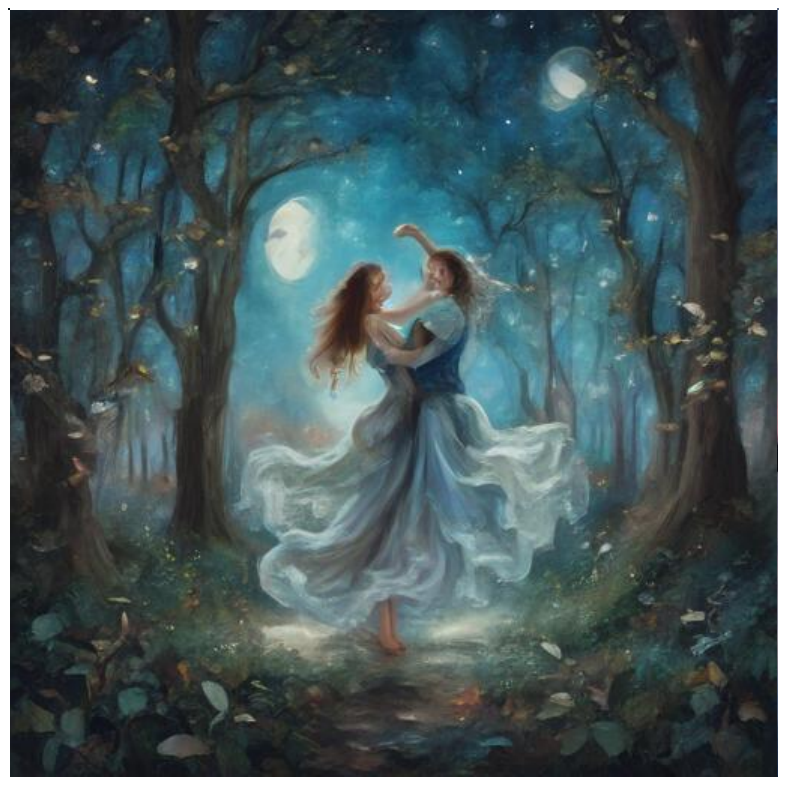}}
	\end{minipage}
   \\
Quantifiable connection - Interstellar &
  \begin{tabular}[c]{p{1\columnwidth}}Twilight over a serene lake, with gentle ripples under a sky painted with soft pastel hues.\end{tabular} &
    \begin{minipage}[b]{0.35\columnwidth}
		\centering
		\raisebox{-.5\height}{\includegraphics[width=\linewidth]{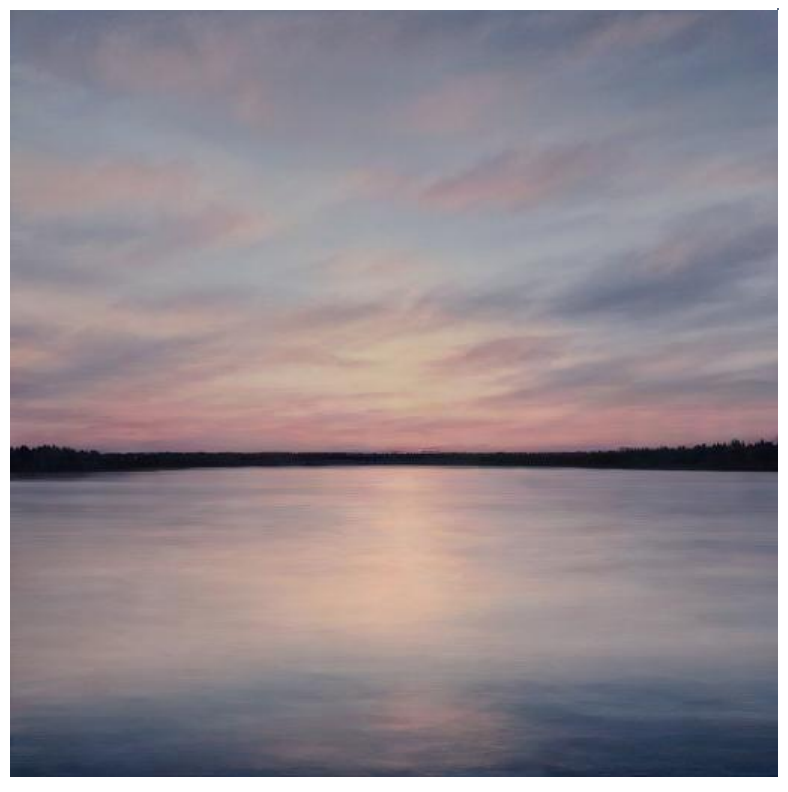}}
	\end{minipage}
   \\
\end{tabular}
}
\caption{Some examples of music clips with corresponding prompts (the description of visual imagery) and visual imagery generated by the system, using the prompt "Based on the perceived emotion of the following musical piece written in ABC notation, and melodic structure/contour, describe a visual imagery people may see when hearing this music. Answer this question only with the text description of the image within 80 token. \{\textit{Music in ABC Notation}\}"}
\label{tab4}
\end{table*}

\subsection{System Architecture and Pipeline}
The system consists of a MIDI keyboard, an LLM-based music analyser, and a visual image generator. 

The MIDI keyboard captures musical inputs as MIDI signals, which are then processed by the system. These signals are buffered and subsequently saved as music clips, which are translated into the text-based ABC Notation format. To build the connection between music and visual imagery, we use emotion as the main medium since visual imagery has been identified as a unique emotional mechanism, a mental projection, of the listener's feelings about the music \citep{emotion_visual1,emotion_visual2}. Our aim is for the generated visual imagery to be emotionally resonant with the input music, enhancing in this way the user's emotional experience during the creative process. Therefore, based on the text-based music input, GPT-4 identifies the perceived emotion, music features (key signature, tempo, harmony, etc), and melodic structure of the input music. This output is then used to describe a visual image that could fit the identified information. The generated output then serves as a prompt for Stable Diffusion XL (SDXL) Turbo, a state-of-the-art performance real-time text-to-image model, to generate the final image. The architecture and workflow of the system are shown in Figure \ref{fig1}. 

Table \ref{tab4} lists some music pieces with the corresponding prompts and visual imagery generated by the system. The stylistic variation of the images is influenced by the genre of the music being played.
For instance, classical pieces typically yield landscapes or atmospheric indoor scenes, while pop and jazz selections inspire the generation of more tangible, realistic scenarios. 


\begin{table}[!ht]
\centering
\setlength{\tabcolsep}{3pt}
\renewcommand{\arraystretch}{1.1}
\begin{tabular}{c|ccccc}
ID & gender & Age & Music & Piano/keyboard &Composition\\ \hline \hline
P1 & M & 29 & $>20$ & 4       & 15          \\ 
P2 & M & 26 & $>20$ & $>16$ & 8           \\ 
P3 & M & 46 & $>20$ & $>20$ & $>20$     \\
P4 & F & 34 & $>20$ & $>20$ & 18          \\ 
P5 & M & 30 & $>20$ & $4$     & $<1$ \\ 
\end{tabular}
\caption{The demographic information and music proficiency of 5 participants in terms of age, gender, music experience, piano/keyboard experience, and composition experience in years.}
\vspace{-0.3cm}
\label{tab5}
\end{table}
\section{Study}
We conducted a user study to explore the feasibility of our system as an inspiration tool to help users create music in an extra-musical cross-modal way, and reveal how it affects user's creative process. In this section, we describe the methodology of the study, the participants, the data collection and the analysis.

\subsection{Methodology}

We designed an experiment to evaluate the system under two music creation scenarios, improvisation and composition. In phase one of the experiment, participants are asked to improvise with the system for ten minutes. In phase two, participants are asked to compose a short music piece of their choice. There are no stylistic or other restrictions on the music. This phase lasted twenty minutes. 

To test user preferences for divergent and convergent inspiration of image generation when using the system in the music creation process,  we designed two versions of the system with different settings of parameter temperature, which is the parameter that controls the degree of randomness in GPT-4's output, with all other factors being identical. After extensive experimentation with different values of temperature (range from 0 to 1), we chose to set the temperatures of divergent and convergent systems to 0.8 and 0.4 respectively, so that the divergent system would not generate many irrelevant images with music, and the convergent system would not keep generating duplicate images. To ensure a fair test of the impact of the two modes of the system, we changed the starting mode among users randomly and switched to the other halfway through the experiment.

After interacting with the system, semi-structured interviews were conducted with each participant to collect qualitative data about their experience with the system.

\subsection{Participants}
We recruited 5 participants, 4 males and 1 female, whose ages ranged from 26 to 46, with a median of 30. Participants were recruited based on the criteria of being musicians with professional training in music, experience in playing the keyboard/piano, and experience in composing at various levels. All participants took part in the experiment voluntarily and received the equivalent of \$30 vouchers for completion of the study.  Basic demographic information about participants is shown in Table \ref{tab5}.

\subsection{Data Collection}
We collected qualitative data using semi-structured interviews. In the interviews, we asked participants about (1) the usual source of inspiration when making music, (2) the emotional association between the system-generated images and the music they played, (3) the impact of the system on music creation, (4) their preference within a divergent generation vs. a convergent generation approach, (5) overall experience with the system, (6) suggestions on improvements. 

\subsection{Data Analysis}
We conducted a thematic analysis of the collected interview data. Two members of the research team coded the interviews' data separately using Nvivo 14, and the final coding was done by integrating and agreeing on the codes between the two coders.

\subsection{Results}
The following section discusses the qualitative results found in the analyses, describing participants' experiences in the experiment.
\subsubsection{Source of Inspiration - Personal Habits of Music Creation:}
Before starting the discussion about the system, the participants were first asked about the usual sources of inspiration in their music creation experience. The purpose of asking this question is to find out if the participants had previous habits of getting inspiration from visual imagery. 

Four of the participants (P2, P3, P4, P5) were usually inspired directly by the music itself; either from the sound of an instrument \say{\textit{More as I just play, and when I hear things that I like, I keep them.}} (P2), \say{\textit{The instrument itself is my source of inspiration}} (P3), \say{\textit{it's just experimenting with sound and seeing the sounds I like}} (P4), or from inspiring melodies \say{\textit{would be some impressive song that I hear recently [...]
Maybe I start from there.}} (P5).

P1, on the other hand, mentioned drawing creative inspiration from narrative rather than music \say{\textit{I guess I'm more interested in the audience experience or the story or the narrative behind the musical experience.}} (P1). P5 reported some external inspirations that did not precede the music creation but happened during the music creation process: \say{\textit{As it's happening, as I'm improvising, [...] 
If at that time I happened to see something like the sunlight coming through the window, I would latch onto that. If it doesn't, then I detach from it.}} (P5). No other participants seemed to consciously use visuals as a source of inspiration. P2 had the experience of composing by looking at pictures when he was trained for composition, but this was not his usual way of composing.

\subsubsection{Influence of Visual Imagery on Music Creativity (RQ1)}
Participants described the impact of the system on their music creation process in a variety of ways. P3 felt that the system made him comfortable with his state of playing. He said, \say{\textit{So I parked my eyes on the images, and then I let my ears conduct my hands. By doing so, my hands started travelling less and just going with the nuances of the attack. And that felt like a good comfort provided by the system}}. P2 and P5 mentioned that the system provides a fluidity of thought in the creative process. This fluidity prevents ideas from getting stuck. As P5 described, \say{\textit{It's just showing something that continuously makes you support your flow, makes your mind move so you don't get stuck [...]
It's a river constantly flowing}}. And P2 said, \say{\textit{Even if the image isn't what I pictured in my head, it still felt fluid.}} P5 described the cross-modal impact that the system has. For example, from visual to kinesthetic, comes from what is in the picture, \say{\textit{when you show the image of a dancer, maybe you think more creatively in terms of playing [...]
maybe you try to enlarge the movement of your finger on the keyboard. Maybe you think about different movements more and emphasize that more}}, and from visual to auditory, \say{\textit{when you show a night sky near the river, maybe that is the difference in the spatial setting, it's a more open space and you tend to create sounds that are more penetrating to the open space.}}

The generated images of our system are based on the perceived emotion of the input music, and this emotional association seems to be an important factor influencing the creative process. Most participants were able to feel this emotional association when using the system except for P1, stating that \say{\textit{they seemed a little bit general}} (P1). P2 felt this emotional association when he started playing with the system, but he stated that \say{\textit{When I kept playing with it, I started to trust it a little bit less. It didn't really diverge when I changed styles}} (P2). The other three participants (P3, P4, P5) thought most of the time they could feel the emotional association. This association influenced the participants' creative process to varying ways and degrees. P4 referred to the effect of this emotional association as the creative atmosphere: \say{\textit{I think it helps create a mood to accompany my creative process, which I guess could keep you in the right work frame of mind. Like it keeps you focused [...] but I don't feel like it directly affects my creation.}} P2 described the effect of this emotional perception as an immersive experience: \say{\textit{I thought it was really cool to have certain images pop up during improvisation because it made me feel like I was sort of there.}} P3 described this emotion as a bell of communicable emotion, a visual dialogue: \say{\textit{when the images that appear with a single character in a big landscape were displayed on the screen, it gave me some sort of meaning that I was only feeling, feeling it at an emotional level. But once they became apparent, once they were displayed on the screen, those feelings were grounded in images. And as such, I think the feeling that I was feeling with the performance was kind of anchored to communicable emotion.}} This visual dialogue is in a sense a guide to him: \say{\textit{at that moment, felt like it was guiding me, not trying, not consciously trying to guide me. But I felt that as soon as I recognized the solitude of those characters, there was a swan of sorts on an empty lake. There was a person on a vast landscape. And my semantic brain started thinking, solitude, loneliness. I accepted it and started playing towards it.}}

It is worth noting that P1 and P2 particularly emphasised the influence of the system in improvisation. They felt that the system provided a good form of improvisation. P1 described the effect of the system in improvisation as \say{\textit{extra material or like a constraint}}, with which he saw the system as a band to work with and communicate with: \say{\textit{I was setting a challenge for myself to try to kind of either match what was being shown or to change what was being shown. And that was kind of fun. It's kind of like if I was improvising within a band, I would kind of do the same kind of dialogue.}}. On the other hand, the task of composing seemed to be disconnected from the system for some participants, with P2 stating that \say{\textit{
It was more like I know what I want this to sound like, and I just got to work out which chord I want here. So I really wasn't looking at the image because I was trying to think about the sound I was making.}}

\subsubsection{Convergent and Divergent Inspiration (RQ2)}
After using both convergent and divergent systems with different degrees of randomness, P1, P3, and P5 perceived that one of the systems generated more repetitive images. P2 and P4 did not perceive a significant difference. Four of the participants agreed that using the divergent system at the beginning of the creative process would be more helpful. P1, P3, and P5 all expressed that there were moments in the creative process when they wished that the images generated by the system would be convergent, for example, P1 stated that \say{\textit{maybe you want it to be less random because you've already kind of landed on this idea. So that can be useful to reinforce something that you're already doing}}, while P5 said that \say{\textit{You want to go into specific directions, maybe you want to pause the system and then refine it.}} P3 provided a valuable insight into this convergent effect: the system should converge on details, not simply repetition: \say{\textit{keep looking at more and more details within the same context, so basically densifying the context as opposed to expanding the context, to explore details.}}

\subsubsection{Enhancing System Experience (RQ3)}
Participants described some of the parts of the system that they would like to improve upon after using the system. P1 felt that he had different needs for the speed at which the system could display images at different stages of his creative process when using the system, \say{\textit{Sometimes I'm like, come on, I want more. And then sometimes, like, wait, just stop on this.}} P2 expected generated images to be more responsive and reactive to what he played during the improvisation, \say{\textit{
I would like it a lot more at moments where I create a very distinct change in what I'm playing for that to be represented within the imagery.}} This timely response was likewise mentioned by P3: \say{\textit{I would love to have pluggable parameters within the image generation that I can map to parameters on the music execution. Say, for example, if I attack more or less, I would love that to be the amount of brightness of these images.}} P3 also extended this response, looking to include more musical features that the player can control to affect the parameters of the image: \say{\textit{
For example, if you get minor chords, you're going to get negative brightness from the previous image. 
That will give me a tool in which if I see something happening image-wise, [...]
this is the forest that I just saw, the lake that I just saw. And I want to make this lake, like a deeper, profound, scary lake. I would learn by ear that these minor chords are giving me that.}} P5 expected more diverse image styles, \say{\textit{
Now it's just like photo-realistic images and very daily pictures. Composition maybe can get inspiration from more abstract things, right? Or maybe just some geometric shapes. 
}}

\section{Discussion}
Our qualitative results reveal that the system influences the creative process in diverse ways, extending beyond emotional resonance to include various cross-modal influences emanating from imagery contents, and vary among individuals. These observed unexpected effects underscore the efficacy of our approach to not impose restrictions on the style and content of the generated images. As a prototype system, there are several improvements to be made. We outline fundamental design considerations derived from our analysis to guide future improvements.
\subsection{Design Consideration}
During the interviews, participants provided valuable insights and suggestions on the use of real-time music-to-image systems as an inspirational tool in the music creation process. Based on these comments, we have summarized and refined them as following design considerations:

\begin{itemize}
    \item \textbf{Variety of Image Style}: Incorporate a diverse range of image styles, including abstract, surreal, and non-photorealistic images, in addition to photorealistic ones. And allow users to customize or control the styles of images generated, as different users may prefer different styles for inspiration.
    \item \textbf{Adaptive Pacing}: Implement adaptive pacing or user-controlled pacing for the displayed of the generated images, to accommodate different preferences and creative needs regarding the timing when the images should be displayed. Adaptions based on musical features such as musical bars and rhythms would be a good way to minimal disruption during the creative process.
    \item \textbf{Responsive Mapping Between Music And Image}: Offer parametric mapping between musical features (e.g., pitch, mode, chord progressions) and visual features (e.g., brightness, colour, motion) to implement a highly responsive and reactive system to distinct changes in user's playing, providing immediate visual feedback within the generated images to build a stronger connection between music and images, opening a more direct creative dialogue.
    \item \textbf{Divergent and Convergent Modes}: Allow users to control the level of divergence or convergence in the generation process. A divergent mode could provide a wide range of diverse images, fostering idea generation and exploration, while a convergent mode could focus on a specific context or theme, providing more detailed and intricate variations within that context, generating coherent, narrative images.
\end{itemize}

\subsection{Limitations and Future Work}
At the core of our prototype system is GPT-4's understanding of music in text-based form. According to our experiments, GPT-4 has \emph{some} text-based music comprehension ability; however, its understanding of complex musical structures and musical contexts is limited. In addition, the text-based music representation format, ABC notation, loses some of the musical information in more complex musical expressions and nuances, such as detailed dynamics, articulations, and ornaments, which can also affect GPT-4's understanding of music. Nonetheless, ABC notation is already the most appropriate form of text-based music representation in terms of meeting the limitations on the token length of GPT-4's prompt. Future improvements to the system may involve fine-tuning GPT models for a better understanding of text-based music representations.

While our user study provides valuable insights into the potential utility of real-time music-to-image systems as creative tools, it is important to acknowledge its limitations. The sample size and participants' diversity are limited, which may constrain the breadth of perspectives and experiences captured. However, it is noteworthy that our participants represent a diverse range of music creation experiences, from novice to professional, encompassing various genres and backgrounds. This diversity allows us to explore the system's applicability across different scenarios during the initial prototype evaluation phase. In the future, we will expand the sample size and conduct a larger study for the final version of the system.

\section{Conclusion}
In this work, we designed the prototype of a real-time music-to-image inspiration system that generates visual imagery based on the emotional expression and musical features of a user's musical input. Our system is based on GPT-4's capabilities for pattern recognition and text-based music understanding.

Through a user study with musicians, we explored the potential of this system as an inspirational tool to enhance creativity in musical improvisation and composition. The key findings indicate that the generated visual imagery resonated emotionally with users to varying degrees and influenced their creative processes in distinct ways, such as creating an immersive atmosphere, providing a fluid source of inspiration, and enabling cross-modal experiences. Most participants reported the system as an engaging and novel tool, particularly valuable for musical improvisation.

Overall, our exploration demonstrates the feasibility of generative AI to facilitate cross-modal inspiration in creative practices like music creation. Future work could involve improving the text-based music understanding capabilities of LLMs, implementing responsive parametric mappings between musical and visual features, and allowing more user control.


\bibliographystyle{iccc}

\end{document}